\documentclass[showpacs,preprint,superscriptaddress]{revtex4-1}
\usepackage{lmodern}

\usepackage[T1]{fontenc}
\usepackage[latin9]{inputenc}
\usepackage[active]{srcltx}
\usepackage{amsmath}
\usepackage{graphicx}
\usepackage{esint}

\makeatletter
\usepackage{amsthm}\usepackage{latexsym}\usepackage{bm}\usepackage{amsfonts}

\makeatother

\begin{document}

\title{Canonical symplectic particle-in-cell method for long-term large-scale
simulations of the Vlasov-Maxwell system}

\author{Hong Qin }

\affiliation{School of Nuclear Science and Technology and Department of Modern
Physics, University of Science and Technology of China, Hefei, Anhui
230026, China}

\affiliation{Plasma Physics Laboratory, Princeton University, Princeton, NJ 08543}

\author{Jian Liu}

\affiliation{School of Nuclear Science and Technology and Department of Modern
Physics, University of Science and Technology of China, Hefei, Anhui
230026, China}

\affiliation{Key Laboratory of Geospace Environment, CAS, Hefei, Anhui 230026,
China}

\author{Jianyuan Xiao}

\affiliation{School of Nuclear Science and Technology and Department of Modern
Physics, University of Science and Technology of China, Hefei, Anhui
230026, China}

\affiliation{Key Laboratory of Geospace Environment, CAS, Hefei, Anhui 230026,
China}

\author{Ruili Zhang}

\affiliation{School of Nuclear Science and Technology and Department of Modern
Physics, University of Science and Technology of China, Hefei, Anhui
230026, China}

\affiliation{Key Laboratory of Geospace Environment, CAS, Hefei, Anhui 230026,
China}

\author{Yang He}

\affiliation{School of Nuclear Science and Technology and Department of Modern
Physics, University of Science and Technology of China, Hefei, Anhui
230026, China}

\affiliation{Key Laboratory of Geospace Environment, CAS, Hefei, Anhui 230026,
China}

\author{Yulei Wang}

\affiliation{School of Nuclear Science and Technology and Department of Modern
Physics, University of Science and Technology of China, Hefei, Anhui
230026, China}

\affiliation{Key Laboratory of Geospace Environment, CAS, Hefei, Anhui 230026,
China}

\author{Yajuan Sun}

\affiliation{LSEC, Academy of Mathematics and Systems Science, Chinese Academy
of Sciences, P.O. Box 2719, Beijing 100190, China }

\author{Joshua W. Burby}

\affiliation{Plasma Physics Laboratory, Princeton University, Princeton, NJ 08543}

\author{Leland Ellison }

\affiliation{Plasma Physics Laboratory, Princeton University, Princeton, NJ 08543}

\author{Yao Zhou}

\affiliation{Plasma Physics Laboratory, Princeton University, Princeton, NJ 08543}
\begin{abstract}
Particle-in-Cell (PIC) simulation is the most important numerical
tool in plasma physics. However, its long-term accuracy has not been
established. To overcome this difficulty, we developed a canonical
symplectic PIC method for the Vlasov-Maxwell system by discretizing
its canonical Poisson bracket. A fast local algorithm to solve the
symplectic implicit time advance is discovered without root searching
or global matrix inversion, enabling applications of the proposed
method to very large-scale plasma simulations with many, e.g., $10^{9}$,
degrees of freedom. The long-term accuracy and fidelity of the algorithm
enables us to numerically confirm Mouhot and Villani's theory and
conjecture on nonlinear Landau damping over several orders of magnitude
using the PIC method, and to calculate the nonlinear evolution of
the reflectivity during the mode conversion process from extraordinary
waves to Bernstein waves. 
\end{abstract}



\maketitle
In modern plasma physics, numerically solving the Vlasov-Maxwell (VM)
equations using the Particle-In-Cell (PIC) method has become the most
important tool \cite{Birdsall91,Hockney88} for theoretical studies
in the last half century. Many innovative algorithms, such as the
Boris scheme for advancing particles \cite{Boris70,Boris77} and Villasenor-Buneman's
charge-conserving deposition scheme \cite{Villasenor92}, have been
developed and successfully applied. Recently, new geometric numerical
methodology has been adopted for PIC simulations. This exciting trend
begins with the discovery of symplectic algorithms for Hamiltonian
equations that govern charged particle dynamics \cite{Ruth83,Feng85,Feng86,Feng10,Forest90,Channell90,Candy91,Marsden01-357,Hairer02,Qin08-PRL,Qin09-PoP}.
The Boris algorithm was discovered to be volume-preserving \cite{Qin13-084503,Zhang15-pop}
and high-order volume-preserving methods have been found \cite{He15}.
In addition, the Vlasov-Maxwell system \cite{Squire12,Xiao13,Kraus14,Shadwick14,Xiao15}
and the Vlasov-Poisson system \cite{Evstatiev13} have been discretized
from a variational symplectic perspective that  preserves symplectic
structures and exhibits excellent long-term accuracy and fidelity. 

In this letter, we develop a new canonical symplectic PIC method for
solving the VM equations by discretizing its canonical Poisson bracket
\cite{Marsden82}. The distribution function $f$ is first discretized
in phase space through the Klimontovich representation by a finite
number of Lagrangian sampling points $(\boldsymbol{\mathbf{X}}_{i},\mathbf{P}_{i})$
$(i=1,...,N)$, where $\mathbf{X}_{i}$ and $\mathbf{P}_{i}$ are
the position and canonical momentum of the $i$-th particle, and $N$
is the total number of sampling points. The electromagnetic field
is discretized point-wise on a given spatial grid, and the Hamiltonian
functional is expressed as a function of the sampling points and the
discretized electromagnetic field. This procedure generates a finite-dimensional
Hamiltonian system with a canonical symplectic structure. The number
of degrees of freedom for the discrete system is $D=3N+3M,$ where
$M$ denotes the total number of the discrete grid-points. 

In general, for a Hamiltonian function whose momentum dependence and
position dependence are not separable, it is not possible to make
symplectic integration algorithms explicit \cite{Feng10,Hairer02}.
For the discrete Hamiltonian system developed here for the VM equations,
the dimension of system is usually very large, and root searching
algorithms required by implicit methods are too time-consuming to
be practical. However, we discovered that if the symplectic Euler
algorithm \cite{Feng10} is applied to the discrete VM Hamiltonian
system at hand, the implicit time advance can be carried out as inexpensively
as an explicit method by just inverting a $3\times3$ matrix for every
particle separately. The resulting canonical symplectic PIC method
for the VM system inherits all the good numerical features of canonical
symplectic algorithms, such as the long-term bound on energy-momentum
error. Being symplectic means that the numerical solution satisfies
$D(2D-1)$ constraints as the exact solution does. Since $D$ is a
large number, the symplectic condition is much stronger than a few
constraints on global energy and momentum. The symplectic condition
is almost as strong as imposing local conservation everywhere in phase
space. 

Two examples of application are given. In the first example, we simulate
the dynamics of nonlinear Landau damping. It also serves as a test
of the algorithm. The discrete VM Hamiltonian system for this study
has more than $2.69\times10^{8}$ degrees of freedom. The damping
rate from the numerical results agrees exactly with the theoretical
result. Furthermore, long-term simulations reveal that the phase mixing
dynamics in velocity space is the physical mechanism of the nonlinear
Landau damping, as recently proved by Mouhot and Villani \cite{Mouhot11,Villani14}
for the Vlasov-Poisson system and conjectured by Villani \cite{Villani14-i}
for the Vlasov-Maxwell system. In the second application, we study
the nonlinear mode conversion process from extraordinary modes to
Bernstein modes (X-B mode conversion) in an inhomogeneous hot plasma.
Simulations show that nonlinear mode excitations and self-interaction
of the Bernstein waves significantly modify the reflectivity and conversion
rate. It is the long-term accuracy and fidelity of the canonical symplectic
PIC algorithm that enables us to numerically confirm Mouhot and Villani's
theory and conjecture over several orders of magnitude using the PIC
method, and to calculate the nonlinear evolution of the reflectivity
during the X-B mode conversion.

We start from the canonical Poission bracket and Hamiltonian for the
Vlasov-Maxwell equations \cite{Marsden82}, 
\begin{eqnarray}
 & \qquad & \{F,G\}\equiv\int f\left\{ \frac{\delta F}{\delta f},\frac{\delta G}{\delta f}\right\} {}_{\mathbf{xp}}d\mathbf{x}d\mathbf{p}+\int\left(\frac{\delta F}{\delta\mathbf{A}}\frac{\delta G}{\delta\mathbf{Y}}-\frac{\delta G}{\delta\mathbf{A}}\frac{\delta F}{\delta\mathbf{Y}}\right)d\mathbf{x}.\label{eq:PB}\\
 & \qquad & H(f,\mathbf{A},\mathbf{Y})=\frac{1}{2}\int(\mathbf{p}-\mathbf{A})^{2}fd\mathbf{x}d\mathbf{p}+\frac{1}{2}\int\left[\mathbf{Y}^{2}+(\nabla\times\mathbf{A})^{2}\right]d\mathbf{x}.\label{eq:H}
\end{eqnarray}
Here, $F$, $G$, and the Hamiltonian $H$ are functionals of the
distribution function $f$, vector potential $\mathbf{A}$, and $\mathbf{Y}\equiv\partial\mathbf{A}/\partial t$.
The bracket $\{h,g\}_{\mathbf{xp}}$ inside the first term on the
right hand side of Eq.\,\eqref{eq:PB} is the canonical Poisson bracket
for functions $h$ and $g$ of canonical phase space $(\mathbf{x},\mathbf{p})$.
The temporal gauge, i.e., $\phi=0$, has been explicitly chosen for
this Poisson bracket to be valid. The Poisson bracket defined in Eq.\,\eqref{eq:PB}
can be formally derived from the point of view of co-adjoint orbit
theory \cite{Marsden82}, and can be used to derive the non-canonical
Morrison-Marsden-Weinstein bracket in the $(f,\mathbf{E},\mathbf{B})$
and $(\mathbf{x},\mathbf{v})$ coordinates \cite{Morrison80,Weinstein81,Marsden82}.
First, we discretize the distribution function using the Klimontovich
representation
\begin{equation}
f(\mathbf{x},\mathbf{p},t)=\sum_{i=1}^{N}\delta\left(\mathbf{x}-\mathbf{X}_{i}\right)\delta\left(\mathbf{p}-\mathbf{P}_{i}\right),
\end{equation}
where $(\mathbf{X}_{i},\mathbf{P}_{i})$ $(i=1,...,N)$ are particles'
coordinates in phase space. Under this discretization, it can be shown
that 
\begin{eqnarray}
\frac{\delta F}{\delta\mathbf{X}_{i}} & = & \int\delta\left(\mathbf{x}-\mathbf{X}_{i}\right)\delta\left(\mathbf{p}-\mathbf{P}_{i}\right)\frac{\partial}{\partial\mathbf{x}}\left(\frac{\delta F}{\delta f}\right)d\mathbf{x}d\mathbf{p},\\
\frac{\delta F}{\delta\mathbf{P}_{i}} & = & \int\delta\left(\mathbf{x}-\mathbf{X}_{i}\right)\delta\left(\mathbf{p}-\mathbf{P}_{i}\right)\frac{\partial}{\partial\mathbf{p}}\left(\frac{\delta F}{\delta f}\right)d\mathbf{x}d\mathbf{p},
\end{eqnarray}
from which we obtain 
\begin{equation}
\frac{\delta F}{\delta\mathbf{X}_{i}}\frac{\delta G}{\delta\mathbf{P}_{i}}-\frac{\delta G}{\delta\mathbf{X}_{i}}\frac{\delta F}{\delta\mathbf{P}_{i}}=\int\delta\left(\mathbf{x}-\mathbf{X}_{i}\right)\delta\left(\mathbf{p}-\mathbf{P}_{i}\right)\left\{ \frac{\delta F}{\delta f},\frac{\delta G}{\delta f}\right\} {}_{\mathbf{xp}}d\mathbf{x}d\mathbf{p}.
\end{equation}
It then follows that the first term on the right-hand side of Eq.\,\eqref{eq:PB}
is
\begin{equation}
\int f\left\{ \frac{\delta F}{\delta f},\frac{\delta G}{\delta f}\right\} {}_{\mathbf{xp}}d\mathbf{x}d\mathbf{p}=\sum_{i=1}^{N}\left(\frac{\delta F}{\delta\mathbf{X}_{i}}\frac{\delta G}{\delta\mathbf{P}_{i}}-\frac{\delta G}{\delta\mathbf{X}_{i}}\frac{\delta F}{\delta\mathbf{P}_{i}}\right).
\end{equation}
Similar derivation of the discretized bracket can be found in the
context of Hamiltonian description of vortex fluid \cite{Morrison81-1783,Morrison81-1788}. 

To discretize the second term on the right-hand side of Eq.\,\eqref{eq:PB},
we first discretize the fields $\mathbf{A}(t)$ and $\mathbf{Y}(t)$
on a Eulerian spatial grid as 
\begin{equation}
\mathbf{A}(\mathbf{x},t)=\sum_{J=1}^{M}\mathbf{A}_{\mathbf{J}}(t)\Psi(\mathbf{x}-\mathbf{x}_{J})\,,\quad\mathbf{Y}(\mathbf{x},t)=\sum_{J=1}^{M}\mathbf{Y}_{J}(t)\Psi(\mathbf{x}-\mathbf{x}_{J})\,,\label{eq:AJ}
\end{equation}
where the discrete fields $\mathbf{A}_{J}(t)$ and $\mathbf{Y}_{J}(t)$
are the fields evaluated on the grid-point $\mathbf{x}_{J}$. The
subscript $J$ is the index of the grid-point, and $M$ is the total
number of the grid-points. Here, $\Psi(\mathbf{x}-\mathbf{x}_{J})$
is the step function, 
\begin{equation}
\Psi(\mathbf{x}-\mathbf{x}_{J})=\left\{ \begin{array}{lc}
1, & |x-x_{J}|<\frac{\Delta x}{2},|y-y_{J}|<\frac{\Delta y}{2},|z-z_{J}|<\frac{\Delta z}{2}~,\\
0, & \textrm{elsewhere}~.
\end{array}\right.
\end{equation}
Under this discretization of $\mathbf{A}(t)$ and $\mathbf{Y}(t)$,
we have
\begin{equation}
\frac{\delta F}{\delta\mathbf{A}}=\sum_{J=1}^{M}\frac{\delta\mathbf{A}_{J}}{\delta\mathbf{A}}\frac{\partial F}{\partial\mathbf{A}_{J}}=\sum_{J=1}^{M}\frac{1}{\Delta V}\Psi(\mathbf{x}-\mathbf{x}_{J})\frac{\partial F}{\partial\mathbf{A}_{J}}\thinspace,\label{eq:dfda}
\end{equation}
where $\Delta V$ is the volume of each cell, which is taken to be
a constant in the present study. For the second equal sign in Eq.\,\eqref{eq:dfda},
use has been made of the fact that 
\begin{equation}
\frac{\delta\mathbf{A}_{J}}{\delta\mathbf{A}}=\frac{1}{\Delta V}\Psi(\mathbf{x}-\mathbf{x}_{J})\thinspace.
\end{equation}
The discretization of the second term on the right-hand side of Eq.\,\eqref{eq:PB}
is thus\emph{ }
\begin{equation}
\int\left(\frac{\delta F}{\delta\mathbf{A}}\frac{\delta G}{\delta\mathbf{Y}}-\frac{\delta G}{\delta\mathbf{A}}\frac{\delta F}{\delta\mathbf{Y}}\right)d\mathbf{x}=\sum_{J=1}^{M}\left(\frac{\partial F}{\partial\mathbf{A}_{J}}\frac{\partial G}{\partial\mathbf{Y}_{J}}-\frac{\partial G}{\partial\mathbf{A}_{J}}\frac{\partial F}{\partial\mathbf{Y}_{J}}\right)\frac{1}{\Delta V}\thinspace.
\end{equation}
Finally, the discrete Poisson bracket for the VM system is 
\begin{equation}
\qquad\left\{ F,G\right\} =\sum_{i=1}^{N}\left(\frac{\delta F}{\delta\mathbf{X}_{i}}\frac{\delta G}{\delta\mathbf{P}_{i}}-\frac{\delta G}{\delta\mathbf{X}_{i}}\frac{\delta F}{\delta\mathbf{P}_{i}}\right)+\sum_{J=1}^{M}\left(\frac{\partial F}{\partial\mathbf{A}_{J}}\frac{\partial G}{\partial\mathbf{Y}_{J}}-\frac{\partial G}{\partial\mathbf{A}_{J}}\frac{\partial F}{\partial\mathbf{Y}_{J}}\right)\frac{1}{\Delta V}\thinspace\label{discPoissonB}
\end{equation}
for functions $F$ and $G$ of the particles $(\mathbf{X}_{i},\mathbf{P}_{i})$
and the discretized field $(\mathbf{A}_{J},\mathbf{Y}_{J})$. 

Next, we need to express the Hamiltonian functional given by Eq.\,\eqref{eq:H}
in terms of $(\mathbf{X}_{i},\mathbf{P}_{i})$ and $(\mathbf{A}_{J},\mathbf{Y}_{J})$.
The particles' total kinetic energy is the sum of each particle's
kinetic energy. The vector potential at a particle's position can
be interpolated from $\mathbf{A}_{J}(t)$ as 
\begin{equation}
\mathbf{A}(\mathbf{X}_{j},t)=\sum_{J=1}^{M}\mathbf{A}_{J}(t)W(\mathbf{X}_{j}-\mathbf{x}_{J})\,,\label{InterpA}
\end{equation}
where $W(\mathbf{\mathbf{X}}_{j}-\mathbf{x}_{J})$ is a chosen interpolation
function. Note that $W(\mathbf{\mathbf{X}}_{j}-\mathbf{x}_{J})$ is
not necessarily the same as the step function $\Psi(\mathbf{x}-\mathbf{x}_{J})$
in Eq.\,\eqref{eq:AJ}. This is of course allowed as long as the
consistency condition is satisfied, i.e., the continuous limit is
recovered when the grid-size goes to zero. The Hamiltonian then becomes

\begin{eqnarray}
 & \qquad & \tilde{H}(\mathbf{X}_{i},\mathbf{P}_{i},\mathbf{A}_{J},\mathbf{Y}_{J})=\frac{1}{2}\sum_{i=1}^{N}\left(\mathbf{P}_{i}^{2}-2\mathbf{P}_{i}\cdot\sum_{J=1}^{M}\mathbf{A}_{J}W(\mathbf{X}_{i}-\mathbf{x}_{J})\right.\nonumber \\
 & \qquad & \left.\sum_{J,L=1}^{M}\mathbf{A}_{J}\cdot\mathbf{A}_{L}W(\mathbf{X}_{i}-\mathbf{x}_{J})W(\mathbf{X}_{i}-\mathbf{x}_{L})\right)+\frac{1}{2}\sum_{J=1}^{M}\left[\mathbf{Y}_{J}^{2}+\left(\nabla_{d}\times\mathbf{A}\right)_{J}^{2}\right]\Delta V,\label{DHamiltonian2}
\end{eqnarray}
where $\left(\nabla_{d}\times\mathbf{A}\right)_{J}$ is the discrete
curl operator acting on the discrete vector potential evaluated at
the $J$-th grid-point. Finally, the discrete Hamiltonian \eqref{DHamiltonian2}
and discrete Poisson structure \eqref{discPoissonB} form a canonical
symplectic discretization of the original continuous Vlasov-Maxwell
system. The ordinary differential equations for the canonical system
are

\begin{eqnarray}
\dot{\mathbf{X}}_{i} & = & \left\{ \mathbf{X}_{i},\tilde{H}\right\} _{d}=\mathbf{P}_{i}-\sum_{J=1}^{M}\mathbf{A}_{J}W(\mathbf{X}_{i}-\mathbf{x}_{J})\,,\label{AMotionEq1}\\
\dot{\mathbf{A}}_{J} & = & \left\{ \mathbf{A}_{J},\tilde{H}\right\} _{d}=\mathbf{Y}_{J}\,,\label{AMotionEq2}\\
\dot{\mathbf{P}}_{i} & = & \left\{ \mathbf{P}_{i},\tilde{H}\right\} _{d}=\sum_{J=1}^{M}\left(\mathbf{P}_{i}\cdot\mathbf{A}_{J}\right)\nabla W(\mathbf{X}_{i}-\mathbf{x}_{J})-\nonumber \\
\qquad &  & \sum_{J,L=1}^{M}\left(\mathbf{A}_{J}\cdot\mathbf{A}_{L}\right)W(\mathbf{X}_{i}-\mathbf{x}_{J})\nabla W(\mathbf{X}_{i}-\mathbf{x}_{L})\,,\label{AMotionEq3}\\
\dot{\mathbf{Y}}_{J} & = & \left\{ \mathbf{Y}_{J},\tilde{H}\right\} _{d}=\sum_{i=1}^{N}\mathbf{P}_{i}W(\mathbf{X}_{i}-\mathbf{x}_{J})\frac{1}{\Delta V}-\nonumber \\
\qquad &  & \sum_{i=1}^{N}\sum_{L=1}^{M}\mathbf{A}_{L}W(\mathbf{X}_{i}-\mathbf{x}_{J})W(\mathbf{X}_{i}-\mathbf{x}_{L})\frac{1}{\Delta V}-\left(\nabla_{d}^{T}\times\nabla_{d}\times\mathbf{A}\right)_{J}\,.\label{AMotionEq4}
\end{eqnarray}
This equation system consists of $6(M+N)$ equations describing the
dynamics of $N$ particles and fields on $M$ discrete grid-points.
The last term in Eq.\,\eqref{AMotionEq4} is defined to be 
\begin{equation}
\left(\nabla_{d}^{T}\times\nabla_{d}\times\mathbf{A}\right)_{J}\equiv\frac{1}{2}\frac{\partial}{\partial A_{J}}\left[\sum_{L=1}^{M}\left(\nabla_{d}\times\mathbf{A}\right)_{L}^{2}\right].\label{eq:dda}
\end{equation}
The notation of $\nabla_{d}^{T}\times\nabla_{d}\times\mathbf{A}$
indicates that the right hand side of Eq.\,\eqref{eq:dda} can be
viewed as the discretized $\nabla\times\nabla\times\mathbf{A}$ for
a well-chosen discrete curl operator $\nabla_{d}$. To wit, we note
that the term $\nabla\times{\bf A}$ in the Hamiltonian is discretized
using the step function $\Psi({\bf x}-{\bf x}_{J})$,
\begin{equation}
\nabla\times{\bf A}=\sum_{J=1}^{M}\left(\nabla_{d}\times{\bf A}\right)_{J}\Psi({\bf x}-{\bf x}_{J}).
\end{equation}
As an example, we define the discrete curl operator $(\nabla_{d}\times\mathbf{A})_{J}=\left(\nabla\times\mathbf{A}\right)_{i,j,k}$
to be 
\begin{equation}
\left(\nabla_{d}\times\mathbf{A}\right)_{J}\equiv\left(\begin{array}{c}
\frac{A_{i,j,k}^{3}-A_{i,j-1,k}^{3}}{\Delta y}-\frac{A_{i,j,k}^{2}-A_{i,j,k-1}^{2}}{\Delta z}\\
\frac{A_{i,j,k}^{1}-A_{i,j,k-1}^{1}}{\Delta z}-\frac{A_{i,j,k}^{3}-A_{i-1,j,k}^{3}}{\Delta{\bf x}}\\
\frac{A_{i,j,k}^{2}-A_{i-1,j,k}^{2}}{\Delta{\bf x}}-\frac{A_{i,j,k}^{1}-A_{i,,j-1,k}^{1}}{\Delta y}
\end{array}\right),
\end{equation}
which can be written as a linear operator on the space of $3M$-vectors
as 
\begin{equation}
\nabla_{d}\times\mathbf{A}=\left(\begin{array}{c}
\left(\nabla_{d}\times\mathbf{A}\right)_{1}\\
\vdots\\
\left(\nabla_{d}\times\mathbf{A}\right)_{M}
\end{array}\right)=\Gamma\left(\begin{array}{c}
\mathbf{A}_{1}\\
\vdots\\
\mathbf{A}_{M}
\end{array}\right)\thinspace.
\end{equation}
Here, $\Gamma$ is a $3M\times3M$ sparse matrix specifying the discrete
curl operator. The partial derivative with respect to $\mathbf{A}_{J}$
can be expressed as 
\begin{equation}
\frac{1}{2}\frac{\partial}{\partial\mathbf{A}_{J}}\left[\sum_{L=1}^{M}(\nabla_{d}\times\mathbf{A})_{L}^{2}\right]=\left[\Gamma^{T}\Gamma\left(\begin{array}{c}
\mathbf{A}_{1}\\
\vdots\\
\mathbf{A}_{M}
\end{array}\right)\right]_{J}\equiv\left(\nabla_{d}^{T}\times\nabla_{d}\times\mathbf{A}\right)_{J},\label{eq:ddaj}
\end{equation}
where $\Gamma^{T}$ is the transposition of $\Gamma$. Obviously,
the notation in Eq.\,\eqref{eq:dda} or \eqref{eq:ddaj} is meaningful
for any linear discrete curl operator $\nabla_{d}$.

It is clear from Eqs.\,\eqref{AMotionEq1}-\eqref{AMotionEq4} that
the particles and fields interact through the interpolation function
$W(\mathbf{X}_{i}-\mathbf{x}_{J})$. The function $W(\mathbf{X}_{i}-\mathbf{x}_{L})/\Delta V$
distributes particles' charge over the grid-points as if they are
``charged clouds'' with finite-size \cite{Birdsall91}.

Once the canonical symplectic structure is given, canonical symplectic
algorithms can be readily constructed using well-developed methods
\cite{Ruth83,Feng85,Feng86,Feng10,Channell90,Forest90,Candy91,Marsden01-357}.
For a reason soon to be clear, we adopt the semi-explicit symplectic
Euler method for time advance. The symplectic Euler method for a generic
canonical Hamiltonian system is
\begin{eqnarray}
p^{n+1} & = & p^{n}-\Delta t\frac{\partial H}{\partial q}(p^{n+1},q^{n}),\\
q^{n+1} & = & q^{n}+\Delta t\frac{\partial H}{\partial p}(p^{n+1},q^{n}).
\end{eqnarray}
where $\Delta t$ is the time-step, and the superscript $n$ in $p^{n}$
and $q^{n}$ denotes that they are the value at the $n$-th time step.
It is implicit for $p$, but explicit for $q$. Making use of this
algorithm, the iteration rules for Eqs.\,(\ref{AMotionEq1}) -(\ref{AMotionEq4})
are

\begin{eqnarray}
 & \qquad & \frac{\mathbf{X}_{i}^{n+1}-\mathbf{X}_{i}^{n}}{\Delta t}=\mathbf{P}_{i}^{n+1}-\sum_{J=1}^{M}\mathbf{A}_{J}^{n}W(\mathbf{X}_{i}^{n}-\mathbf{x}_{J})\,,\label{BMotionEq1}\\
 & \qquad & \frac{\mathbf{A}_{J}^{n+1}-\mathbf{A}_{J}^{n}}{\Delta t}=\mathbf{Y}_{J}^{n+1}\,,\label{BMotionEq2}\\
 & \qquad & \frac{\mathbf{P}_{i}^{n+1}-\mathbf{P}_{i}^{n}}{\Delta t}=\sum_{J=1}^{M}\left(\mathbf{P}_{i}^{n+1}\cdot\mathbf{A}_{J}^{n}\right)\nabla W(\mathbf{X}_{i}^{n}-\mathbf{x}_{J})\nonumber \\
 & \qquad & \qquad-\sum_{J,L=1}^{M}\left(\mathbf{A}_{J}^{n}\cdot\mathbf{A}_{L}^{n}\right)W(\mathbf{X}_{i}^{n}-\mathbf{x}_{J})\nabla W(\mathbf{X}_{i}^{n}-\mathbf{x}_{L})\,,\label{BMotionEq3}\\
 & \qquad & \frac{\mathbf{Y}_{J}^{n+1}-\mathbf{Y}_{J}^{n}}{\Delta t}=\sum_{i=1}^{N}\mathbf{P}_{i}^{n+1}W(\mathbf{X}_{i}^{n}-\mathbf{x}_{J})\frac{1}{\Delta V}\nonumber \\
 & \qquad & \qquad-\sum_{i=1}^{N}\sum_{L=1}^{M}\mathbf{A}_{L}^{n}W(\mathbf{X}_{i}^{n}-\mathbf{x}_{J})W(\mathbf{X}_{i}^{n}-\mathbf{x}_{L})\frac{1}{\Delta V}-\left(\nabla_{d}^{T}\times\nabla_{d}\times\mathbf{A}^{n}\,\right)_{J}.\label{BMotionEq4}
\end{eqnarray}
These difference equations furnish a canonical symplectic PIC method
for the Vlasov-Maxwell equations. 

As discussed above, symplectic algorithms for a Hamiltonian system
with non-separable momentum and position dependence are implicit in
general. This is indeed the case for the difference equations \,\eqref{BMotionEq1}-\eqref{BMotionEq4},
because the right-hand sides of Eqs.\,\eqref{BMotionEq1}-\eqref{BMotionEq4}
depend on values of the $(n+1)$-th time-step. However, they are semi-explicit,
because Eqs.\,\eqref{BMotionEq1}, \eqref{BMotionEq2}, and \eqref{BMotionEq4}
are explicit for $\mathbf{X}_{i}^{n+1}$, $\mathbf{A}_{J}^{n+1}$,
and $\mathbf{Y}_{J}^{n+1},$ respectively. Another good property of
the system is that the only implicit equation \eqref{BMotionEq3}
is linear in terms of $\mathbf{P}_{i}^{n+1}$, and it is only implicit
for each particle, i.e., Eq.\,\eqref{BMotionEq3} does not couple
$\mathbf{P}_{i}^{n+1}$ and $\mathbf{P}_{k}^{n+1}$ when $i\neq k.$
Therefore, the system can be solved without root searching iterations
as follows. We first solve the linear equation \eqref{BMotionEq3}
for $\mathbf{P}_{i}^{n+1}$ for every index $i$ separately, which
amounts to inverting a $3\times3$ matrix for every $i$. Then $\mathbf{X}_{i}^{n+1}$
and $\mathbf{Y}_{J}^{n+1}$ are advanced explicitly according to Eqs.\,\eqref{BMotionEq1}
and \eqref{BMotionEq4}, and the last step is to advance $\mathbf{A}_{J}^{n+1}$,
also explicitly, according to Eq.\,\eqref{BMotionEq2}. 

The preservation of the symplectic structure exerts $D(2D-1)$ constraints
on the numerical solution. Because $D$ is a large number, preservation
of symplectic structure is a very strong constraint and significantly
reduces the errors of numerical solutions. We can also appreciate
this advantage from the viewpoint of symplectic capacity, which is
defined on any open set of the phase space. Symplectic maps preserves
symplectic capacity \cite{Hofer94}, and in principle there are infinite
constraints that symplectic algorithms can satisfy as the continuous
systems do.

\begin{figure}
\centering{}\includegraphics[width=9cm]{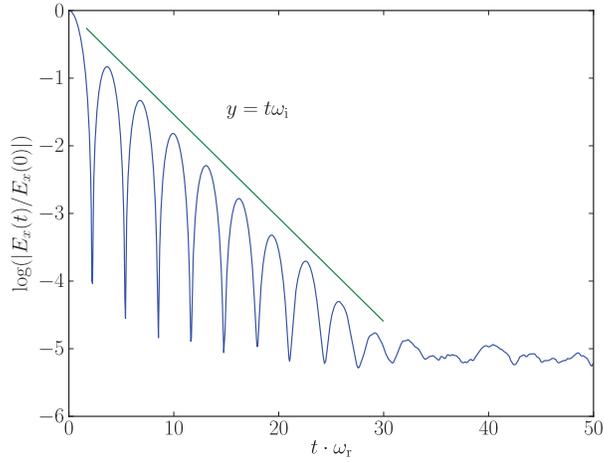}\caption{Perturbed electrical field as a function of time. The slope of the
green line is the theoretical damping rate.\label{fig1}}
\end{figure}

\begin{figure}
\begin{centering}
\includegraphics[width=9cm]{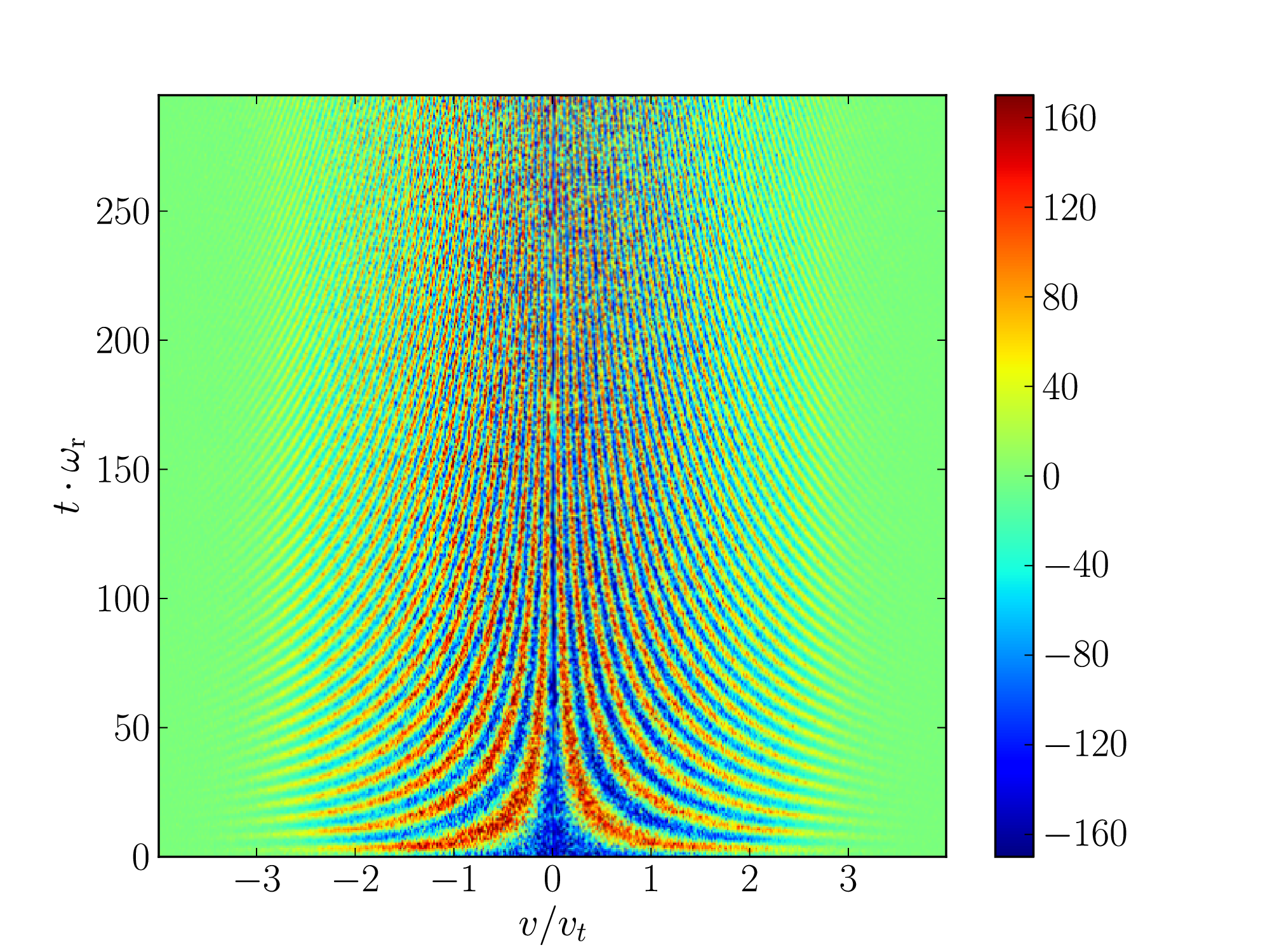}
\par\end{centering}

\caption{Electron distribution function in velocity space as a function of
time. Different colors denote the amplitude of the perturbation. The
mechanism of phase mixing in velocity space is clearly demonstrated.
The wave-number in velocity space increases with time, which results
in a decrease in density perturbation and thus attenuation of the
electrical field.\label{fig2}}
\end{figure}

We now apply this canonical symplectic PIC scheme to simulate the
nonlinear Landau damping process. This study also serves as a test
of the algorithm. Previously, similar study and test have been performed
for other algorithms, e.g., the Eulerian algorithms for the Vlasov-Poisson
system \cite{Zhou01,Kraus14}. The ions are treated as a uniform positively
charged background, and the dynamics of electrons are simulated. The
electron density is $n_{e}=1.149\times10^{16}/m^{3}$, and the thermal
velocity of electrons is $v_{T}=0.1c$, where $c$ is the light velocity
in vacuum. The three-dimensional computational region is divided into
$896\times1\times1$ cubic cells. The size of grid is chosen to be
$\Delta x=2.4355\times10^{-4}m$, the time step $\Delta t=\Delta x/2c$.
The interpolation function is chosen to be 8th order, i.e.,
\begin{eqnarray}
W(\mathbf{x}) & = & W_{1}(x/\Delta x)W_{1}(y/\Delta x)W_{1}(z/\Delta x)~,\\
W_{1}(q) & = & \left\{ \begin{array}{lc}
0, & q>2~,\\
\frac{15}{1024}q^{8}-\frac{15}{128}q^{7}+\frac{49}{128}q^{6}-\frac{21}{32}q^{5}+\frac{35}{64}q^{4}-q+1, & 1<q\leq2~,\\
-\frac{15}{1024}q^{8}-\frac{15}{128}q^{7}+\frac{7}{16}q^{6}-\frac{21}{32}q^{5}+\frac{175}{256}q^{4}-\frac{105}{128}q^{2}+\frac{337}{512}, & 0<q\leq1~,\\
-\frac{15}{1024}q^{8}+\frac{15}{128}q^{7}+\frac{7}{16}q^{6}+\frac{21}{32}q^{5}+\frac{175}{256}q^{4}-\frac{105}{128}q^{2}+\frac{337}{512}, & -1<q\leq0~,\\
\frac{15}{1024}q^{8}+\frac{15}{128}q^{7}+\frac{49}{128}q^{6}+\frac{21}{32}q^{5}+\frac{35}{64}q^{4}+q+1, & -2<q\leq-1~,\\
0, & q<-2~.
\end{array}\right.
\end{eqnarray}
It can be proved that the kernel function $W$ is 3rd order continuous
in the whole space. According to our performance benchmarks, this
8th order kernel is about 30\% more computationally costly than a
2nd order kernel, which is acceptable since a higher order continuous
kernel gives more numerical fidelity. Initially, $10^{5}$ sampling
points of electrons are distributed in each cell. The total number
of particles is $N=8.96\times10^{7}$, and the number of degrees of
freedom is $D=2.69\times10^{8}.$ The initial electrical field perturbation
is $\mathbf{E}_{1}=E_{1}\cos\left(kx\right)\mathbf{e}_{x},$ where
the wave number is $k=2\pi/224\Delta x$, and the amplitude of the
perturbation electric field is $E_{1}=9.103\times10^{4}V/m.$ The
simulations are performed for 80000 time steps, during which a complete
picture of the nonlinear Landau damping is revealed. As expected for
symplectic algorithms, the numerical error of energy does not increase
with time and is bounded within 1\% for all time. The theoretical
damping rate calculated from the dispersion relation is $\omega_{i}=-1.3926\times10^{9}/s$,
and the theoretical real frequency is $\omega_{r}=9.116\times10^{9}/s$.
In Fig.\ \ref{fig1}, the slope of the green line is the theoretical
damping rate, and the blue curve is the evolution of the electrical
field observed in the simulation. It is evident that the simulation
and theory agree perfectly. After $t=30/\omega_{r}$, the energy of
the wave drops below the level of numerical noise, and the damping
process stops. The evolution of the electron distribution function
is plotted in Fig.\ \ref{fig2}, which clearly demonstrates the mechanism
of phase mixing in velocity space. We observe in Fig.\ \ref{fig2}
that the wave-number in velocity space increases with time, which
results in a decrease in density perturbation and thus attenuation
of the electrical field. More importantly, this mechanism of phase
mixing is the dominant physics for the entire nonlinear evolution
of the Landau damping, as proved by Mouhot and Villani \cite{Mouhot11,Villani14}
recently for the electrostatic Vlasov-Poisson system. In addition,
our simulation is electromagnetic and it shows that this physical
picture of nonlinear Landau damping is also valid for the electromagnetic
Vlasov-Maxwell system, as Villani conjectured \cite{Villani14-i}.
It is the long-term accuracy and fidelity of the canonical symplectic
PIC algorithm that enables the confirmation of Mouhot and Villani's
theory and conjecture over several orders of magnitude.

\begin{figure}
\begin{centering}
\includegraphics[width=9cm]{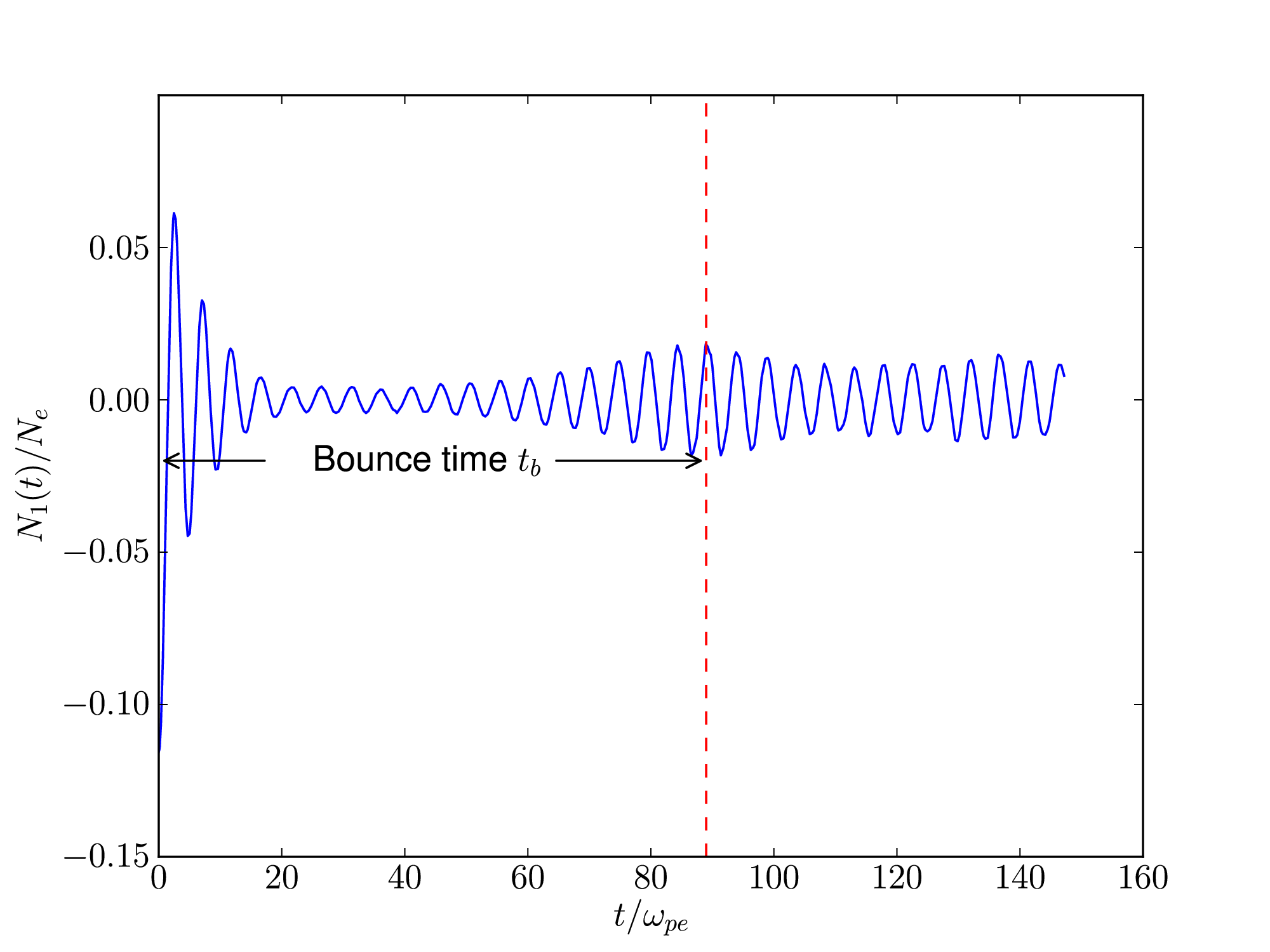}
\par\end{centering}

\caption{In nonlinear Landau damping, the perturbation will bounce back after
initial phase of damping if the initial perturbation is large enough.
The amplitude of the initial electric field in this case is 0.494MV/m.
\label{fig3}}
\end{figure}

\begin{figure}
\begin{centering}
\includegraphics[width=9cm]{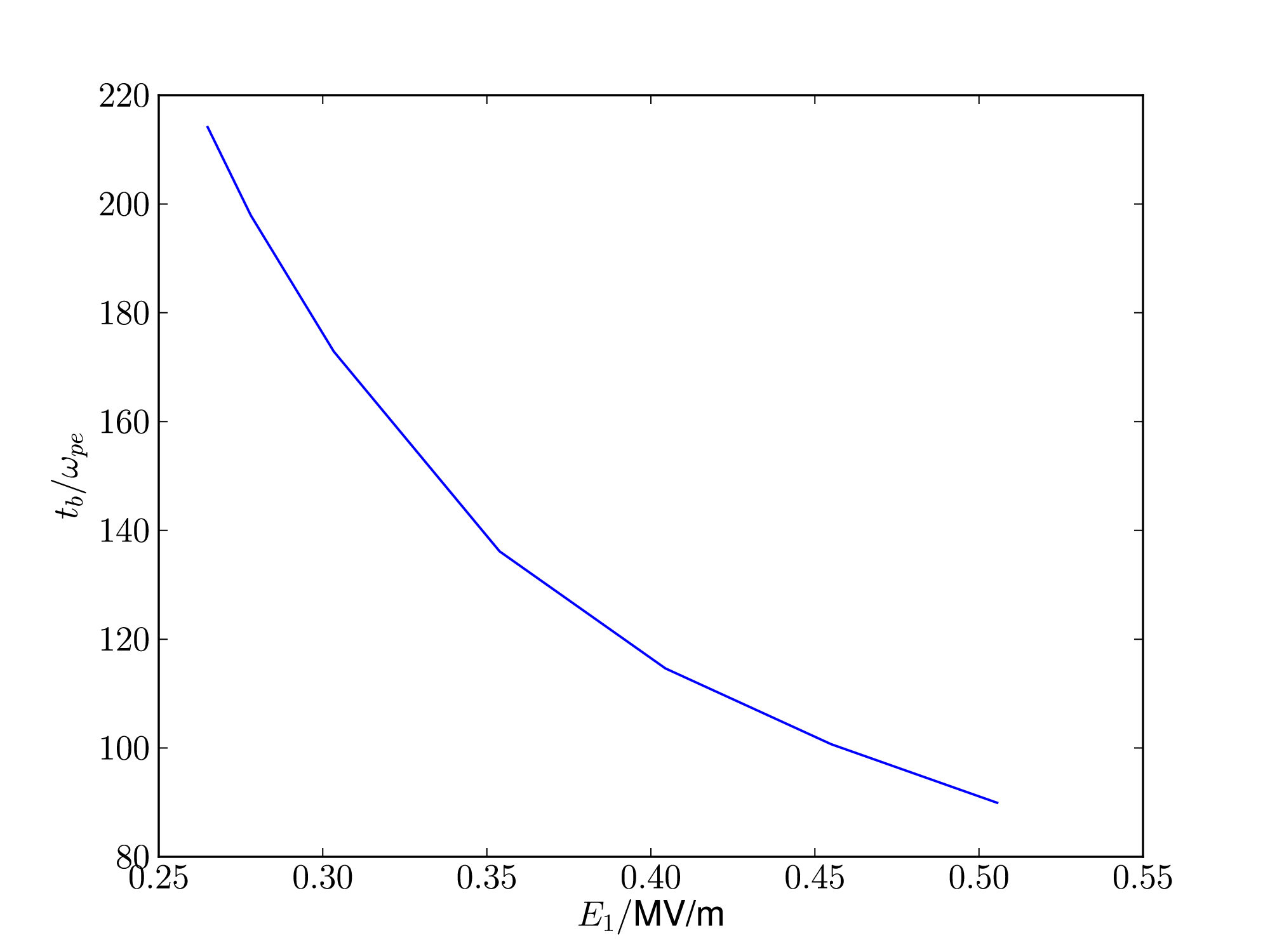}
\par\end{centering}

\caption{Relationship between the initial amplitude $E_{1}$ and the bounce
time $t_{b}$ in nonlinear Landau damping. \label{fig4}}
\end{figure}

Even though Mouhot and Villani rigorously proved that when the initial
perturbation amplitude is small enough, the electrical perturbation
will decay to zero, plasma physicists have known this fact since 1960s
\cite{Armstrong67,Canosa74}. It has also been known that when the
initial electrical perturbation is large enough, the perturbation
will bounce back after initial phase of damping \cite{Armstrong67,Canosa74}.
One such case simulated is plotted in Fig.\,\ref{fig3}. For this
case, the amplitude of the initial electrical field is 0.494MV/m and
the wave number is $k=2\pi/272\Delta x$. Figure \ref{fig4} shows
the bounce-time as a function of the initial amplitude of the electrical
field obtained in simulations. The physics demonstrated in our PIC
simulations agrees with that obtained from Eulerian solvers \cite{Zhou01},
except that our simulations are carried out for the full Vlasov-Maxwell
system.


\begin{figure}
\centering{}\includegraphics[width=10cm]{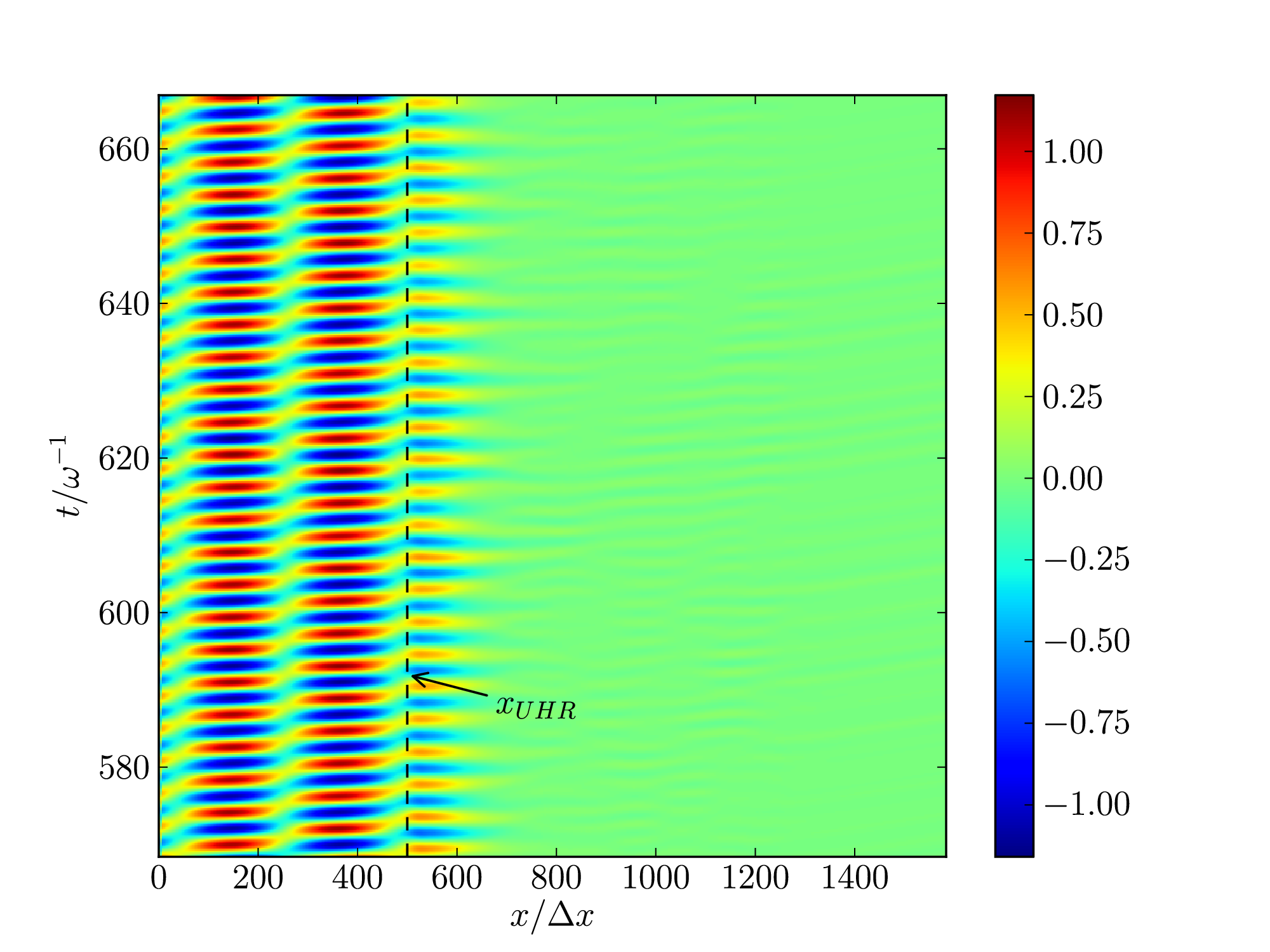}\caption{The space-time dependence of $E_{y}(t,x)$ during the nonlinear X-B
mode conversion.}
\label{fig5}
\end{figure}

\begin{figure}
\centering{}\includegraphics[width=10cm]{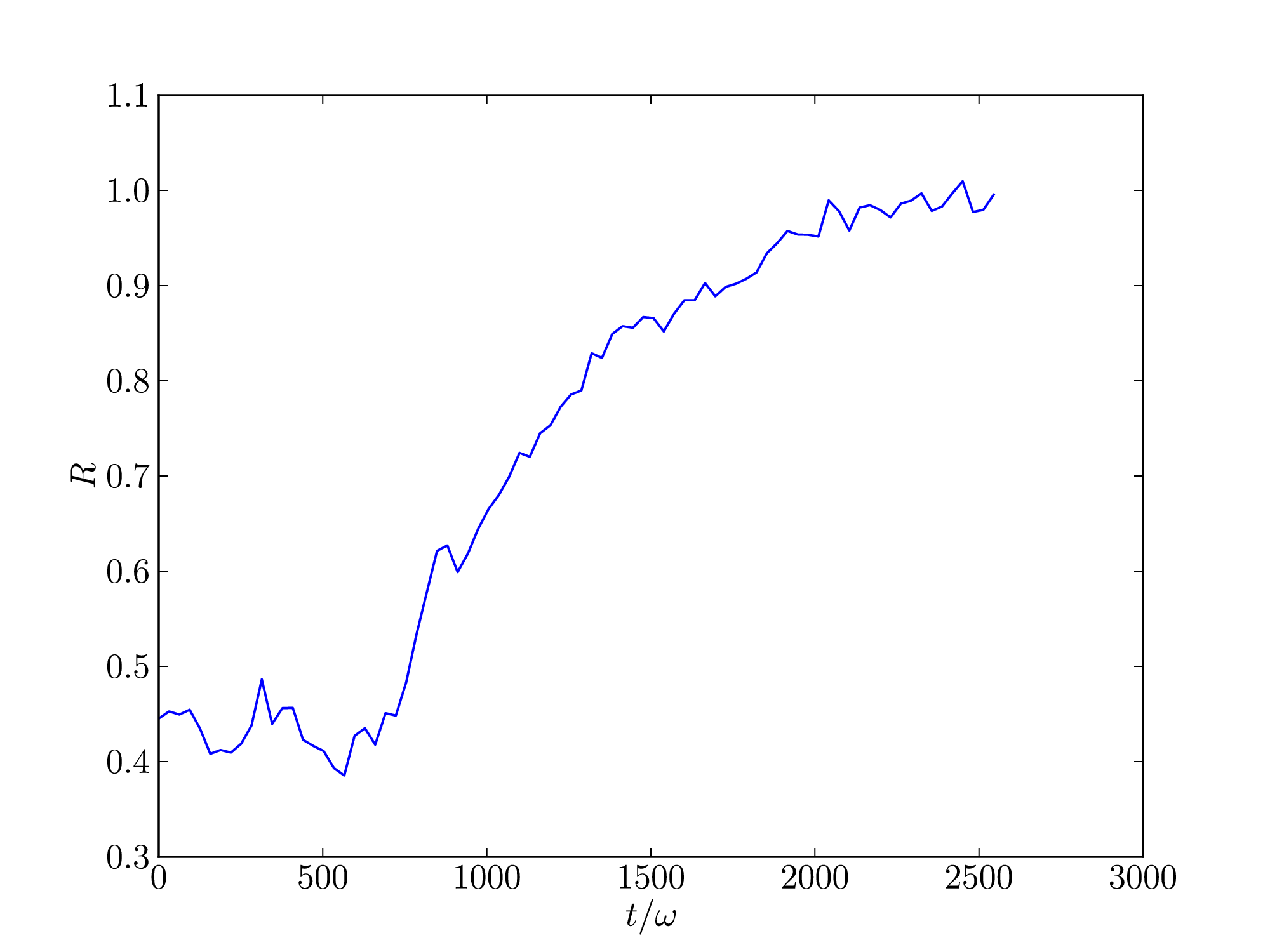}\caption{The evolution of reflectivity during the nonlinear X-B mode conversion.}
\label{fig6}
\end{figure}

In the second application, the 1D nonlinear mode conversion of extraordinary
waves to electron Bernstein waves (X-B mode conversion) in a inhomogeneous
hot plasma is simulated for a long time. The plasma density profile
in the simulation is \\
\begin{equation}
n_{e}(x)=n_{0}\left\{ \begin{array}{lc}
\exp\left[-\left(\frac{x/\Delta x-n_{r}-320}{0.4n_{r}}\right)^{2}\right]~, & 0<\frac{x}{\Delta x}\leq\left(n_{r}+320\right)~,\\
1~, & \left(n_{r}+320\right)<\frac{x}{\Delta x}\leq1300~,
\end{array}\right.
\end{equation}
where $n_{0}=2.3\times10^{19}~\mathrm{m}^{-3},$ $n_{r}=380,$ and
$\Delta x=2.773\times10^{-5}$m is the grid size. The thickness of
the plasma boundary is $n_{r}\Delta x$. The electron temperature
is $T_{e}=57.6\mathrm{{eV}}$, and constant external magnetic field
is $\mathbf{B}=B_{0}\mathbf{e}_{z}$ with $B_{0}=0.6\mathrm{{T}}$.
The simulation domain is a $1584\times1\times1$ cubic mesh. At both
boundaries in the $x$-direction, the Mur's absorbing condition are
used, and periodic boundary conditions are adopted in the $y$- and
$z$-directions. The time step is chosen to be $\Delta t=\Delta x/2c$.
At the left boundary, a source is placed to excite an electromagnetic
perturbation at $\omega=0.0145/\Delta t$ with $\mathbf{E}_{1}=E_{1}\mathbf{e}_{y}$
and $E_{1}=900\mathrm{{kV}}$. As illustrated in Fig.\,\ref{fig5},
the extraordinary wave excited at the left boundary first propagates
to the region of cutoff-resonance-cutoff near $x/\Delta x=500.$ The
wave is then partially reflected back, and partially converted to
electron Bernstein waves \cite{Xiao15}. The reflectivity evolution
is plotted in Fig. \ref{fig6}. Nonlinear excitations and self-interactions
of the Bernstein modes dominate the long time dynamics of the mode
conversion process. As a consequence, the reflectivity of the incident
wave evolves nonlinearly as shown in Fig.\,\ref{fig6}. For this
set of chosen parameters, the incident wave is completely reflected
at the later time of the process. 

In conclusion, we have developed a canonical symplectic particle-in-cell
simulation method for the Vlasov-Maxwell system by discretizing its
canonical Poisson bracket. In phase space, the distribution function
is discretized by the Klimontovich representation using Lagrangian
markers, and the electromagnetic field is discretized point-wise on
a spatial grid. The resulting canonical Hamiltonian system with a
large number of degrees of freedom is integrated by the symplectic
Euler method, whose difference equations can be solved inexpensively
by inverting a $3\times3$ matrix locally for every particle. Implicit
root searching and global matrix inversion are avoided entirely. This
technique makes large-scale applications of the developed canonical
symplectic PIC method possible. To suppress numerical noise caused
by the coarse sampling, smoothing functions for sampling points can
also be conveniently implemented in the canonical symplectic PIC algorithm.
By incorporating the smoothing functions into the Hamiltonian functional
before the discretization, we are able to rein in all the benefits
of smoothing functions without destroying the canonical symplectic
structure. Progress in this and other directions will be reported
in future publications. 
\begin{acknowledgments}
This research is supported by ITER-China Program (2015GB111003, 2014GB124005,
2013GB111000), JSPS-NRF-NSFC A3 Foresight Program in the field of
Plasma Physics (NSFC-11261140328), the CAS Program for Interdisciplinary
Collaboration Team, the Geo-Algorithmic Plasma Simulator (GAPS) project,
and the the U.S. Department of Energy (DE-AC02-09CH11466).
\end{acknowledgments}


\end{document}